\documentclass[aps,twocolumn,prd,superscriptaddress,preprintnumbers]{revtex4}
\usepackage{graphicx}

\newcommand{\be}{\begin{equation}}
\newcommand{\ee}{\end{equation}}
\newcommand{\bq}{\begin{eqnarray}}
\newcommand{\eq}{\end{eqnarray}}
\newcommand{\bsq}{\begin{subequations}}
\newcommand{\esq}{\end{subequations}}
\newcommand{\bc}{\begin{center}}
\newcommand{\ec}{\end{center}}

\newcommand\lsim{\mathrel{\rlap{\lower4pt\hbox{\hskip1pt$\sim$}}
    \raise1pt\hbox{$<$}}}
\newcommand\gsim{\mathrel{\rlap{\lower4pt\hbox{\hskip1pt$\sim$}}
    \raise1pt\hbox{$>$}}}
\begin{document}

\preprint{FTPI-MINN-06/16}
\preprint{UMN-TH-2503/06}

\title{Reconstructing the dark energy equation of state with varying couplings}

\author{P. P. Avelino}
\email[Electronic address: ]{ppavelin@fc.up.pt}
\affiliation{Centro de F\'{\i}sica do Porto, Rua do Campo Alegre 687, 4169-007 Porto, Portugal}
\affiliation{Departamento de F\'{\i}sica da Faculdade de Ci\^encias
da Universidade do Porto, Rua do Campo Alegre 687, 4169-007 Porto, Portugal}
\author{C. J. A. P. Martins}
\email[Electronic address: ]{C.J.A.P.Martins@damtp.cam.ac.uk}
\affiliation{Centro de F\'{\i}sica do Porto, Rua do Campo Alegre 687, 4169-007 Porto, Portugal}
\affiliation{Department of Applied Mathematics and Theoretical Physics,
Centre for Mathematical Sciences,\\ University of Cambridge,
Wilberforce Road, Cambridge CB3 0WA, United Kingdom}
\author{N. J. Nunes}
\email[Electronic address: ]{nunes@physics.umn.edu}
\affiliation{School of Physics and Astronomy, University of Minnesota, 116 Church Street S.E., Minneapolis, Minnesota 55455, USA}
\author{K. A. Olive}
\email[Electronic address: ]{olive@physics.umn.edu}
\affiliation{School of Physics and Astronomy, University of Minnesota, 116 Church Street S.E., Minneapolis, Minnesota 55455, USA}

\date{29 August 2006}

\begin{abstract}
We revisit the idea of using varying couplings to probe the nature of dark energy, in particular by reconstructing its equation of state. We show that for the class of models studied this method can be far superior to the standard methods (using type Ia supernovae or weak lensing). We also show that the simultaneous use of measurements of the fine-structure constant $\alpha$ and the electron-to-proton mass ratio $\mu$ allows a direct probe of grand unification scenarios. We present forecasts for the sensitivity of this method, both for the near future and for the next generation of spectrographs --- for the latter we focus on the planned CODEX instrument for ESO's Extremely Large Telescope (formerly known as OWL). A high-accuracy reconstruction of the equation of state may be possible all the way up to redshift $z\sim4$.
\end{abstract}

\pacs{98.80.Es, 95.36.+x, 98.80.Cq}
\maketitle

\section{\label{intro}Introduction}

Observations suggest that the recent universe is dominated by an energy component whose gravitational behaviour is quite similar to that of a cosmological constant (as first introduced by Einstein) \cite{Riess,Eisenstein,Spergel}. This could of course turn out to be the cause for the recent acceleration of the Universe, but the observationally required value is much smaller than what would be expected from particle physics. In addition to the magnitude of the vacuum energy, we are also faced with the cosmic coincidence that the vacuum energy density is of the same order of magnitude as the matter density. A dynamical scalar field is arguably a plausible explanation of the latter problem \cite{Copeland}. Theoretical motivation for such a field is not hard to find. In string theory, for example, dimensional parameters are expressed in terms of the string mass scale and a scalar field vacuum expectation value. 

A crucial observational goal is therefore to characterize the properties of dark energy, and in particular to look for evidence of dynamical behaviour. A simple but important property is its equation of state, $w=p/\rho$, and considerable effort has recently been put into trying to measure it as a function of redshift. The current methods of choice are type Ia supernovae and (more recently) weak lensing, which is slightly more promising than the former. However, the question arises as to whether these are indeed the best tools for the task at hand. It has been known for some time \cite{Maor1,Maor2} that supernova measurements are limited as a probe of the dark energy equation of state, especially if it is varying with redshift. A discussion of current and future constraints on the dark energy equation of state, from the various standard approaches and parametrized in the usual way, is given in Ref.~\cite{Upadhye}, and it is easy to conclude from it that a convincing detection of time variation of $w$ is unlikely even with hypothetical future space-based experiments such as DUNE or JDEM (be it in its SNAP or DESTINY incarnations). This is not surprising since any dynamical field providing the dark energy must be slow-rolling at the present time (this is mandatory in order to have acceleration), and for slow variations there will always be a constant $w$ model that produces nearly identical results over the redshift range where dark energy is dynamically important.

So, it is important to ask whether better (and cheaper) alternatives are available. A potentially effective  tool for probing dynamical dark energy has been suggested previously in \cite{Parkinson,Nunes,Doran}, though not yet studied in detail. In any realistic model where the dark energy is due to a dynamical scalar field, that same field is also expected to produce sizable varying couplings \cite{Wett,Wett2,Carroll}. While some phenomenological studies of these models have been carried out 
\cite{Livio,Landau,Chamoun,Sandvik,Pospelov,Chiba,Anchordoqui,Nunes0,Bento,Oliveira,Marra,Avelino,Lee1,Byrne,Fujii,Lee2}, 
we show that probing these couplings can be a key test of these class of models, and in particular the varying couplings can be used to infer the evolution of the scalar field, and thus determine its equation of state. 
This is analogous to reconstructing the 1D potential for the classical motion of a particle once its trajectory has been specified.

Previous efforts \cite{Parkinson,Nunes,Doran} only considered the variation of the fine-structure constant $\alpha$, for which there is now a considerable dataset of observations and some (albeit disputed) evidence for variations in the redshift range $z\sim1-3$ \cite{Webb,Murphy,Chand,Srianand}---see also \cite{Uzan,Phil} for reviews. However, in theories where a dynamical scalar field is responsible for varying $\alpha$, the other gauge and Yukawa couplings are also expected to vary \cite{Campbell}. Specifically, in GUTs there is a relation between the variation of $\alpha$ and that of the QCD scale, $\Lambda_{QCD}$, implying that the nucleon mass will vary when measured in units of the Planck mass.
Similarly, one would expect variations in the Higgs vacuum expectation value (VEV), $v$, leading to changes in all particle mass scales including the electron mass. We therefore expect variations of the proton-to-electron mass ratio, $\mu=m_p/m_e$. Some specific models are discussed in Refs.~\cite{Campbell,Calmet1,Calmet2,Langacker,Olive,Dine}. Typically the relation between the variations is
\be
\frac{\dot\mu}{\mu}\sim\frac{\dot\Lambda_{QCD}}{\Lambda_{QCD}} - \frac{\dot{v}}{v} 
\sim R\frac{\dot \alpha}{\alpha};\label{defR}
\ee
the latter equality should be seen as the first term in a Taylor series, but given the expected level of variations the approximation should be good enough for most purposes. The value of $R$ is model-dependent (indeed, even its sign is not determined \textit{a priori}), but large values 
and negative values are generic for GUT models in which modifications come from high-energy scales:
typically $\dot\Lambda/\Lambda \sim 30 \dot\alpha/\alpha$ and $\dot{v}/v \sim 80 \dot\alpha/\alpha$
so that $R \sim -50$.  The large proportionality factors arise simply because the strong coupling constant and the Higgs VEV run (exponentially) faster than $\alpha$.

If this intuitive picture is correct, then variations of $\mu$ may be easier to detect than those of $\alpha$. As first pointed out in Ref.~\cite{Thompson}, astrophysical measurements of $\mu$ can be made using $H_2$ molecular lines. Some recent measurements using this technique \cite{Ivanchik,Ubachs} also find evidence for a variation of $\mu$, although the number of current measurements is currently much smaller than the $\alpha$ dataset---the main reason for this is the difficulty in finding molecular Hydrogen clouds. One of the goals of the present paper is to encourage further measurements of $\mu$, which are expected to lead to tighter constraints on the evolution of the dark energy equation of state than those of $\alpha$ if $R$ is indeed large.

\section{\label{theory}Reconstructing the equation of state}

As we already pointed out, it was claimed that one could in principle extract information on the evolution of the equation of state of dark energy through the redshift dependence of $\alpha$ measurements \cite{Parkinson,Nunes,Doran}. Here we briefly review the reconstruction procedure and show that including measurements of $\mu$ in this approach allows us to estimate the value of the parameter $R$. We will concentrate here on minimally coupled models of dark energy.

The reconstruction pipeline follows the procedure introduced in Ref.~\cite{Nunes} for $\alpha$. We assume that the functional form of the variation of these constants can be simply parametrized by a polynomial such that 
\begin{equation}
\label{defg}
g_x(N) \equiv \frac{\Delta x}{x} = {g_x}_1 N + {g_x}_2 N^2 + ... + {g_x}_m N^m \,,
\end{equation}
where the various ${g_x}_i$ are the coefficients of the polynomial, $N = -\ln (1+z)$ and $x$ stands for either $\alpha$ or $\mu$. Given the tight Weak Equivalence Principle bounds \cite{Livio,Pospelov}, we also assume that the evolution of $\alpha$ and $\mu$ depends linearly on the value of the quintessence field through 
\begin{equation}
\label{lin}
\frac{\Delta x}{x} = \zeta_x \kappa (\phi - \phi_0) \,,
\end{equation}
and that $\zeta_\mu = R \zeta_\alpha$ in agreement with Eq.~(\ref{defR}).

Under these assumptions it was shown in Ref~\cite{Nunes} that the equation of motion for the energy density of the scalar field can be written as
\begin{equation}
\label{eqsigma}
\sigma' = -\left(\frac{g_x'}{\zeta_x}\right)^2 \, \left(1 + a^{-3}\right)\,,
\end{equation}
where $\sigma = \rho_\phi/\rho_0 \Omega_{M0}$ and a prime denotes differentiation with respect to $N$. The equation of state is obtained from the relation $w = -1 + \dot{\phi}^2/\rho_\phi$ which can be rewritten as:
\begin{equation}
\label{eqw}
w(N) = -1+ \frac{1}{3} \left(\frac{g_x'}{\zeta_x}\right)^2 \, \left(1 + \frac{1}{\sigma a^3}\right) \,.
\end{equation}
Thus the evolution of the equation of state is determined by fitting the polynomial (\ref{defg}) to the data, solving the differential equation (\ref{eqsigma}) and by using its result in (\ref{eqw}).

The reconstruction under this procedure is ambiguous regarding the overall scale since the value of $\zeta_x$ is unknown. Hence a normalization of the equation of state must be performed by fixing its value at a given redshift to a value compatible with the value measured via independent observations. More specifically, knowing the present values $w_0$ and $\Omega_{\phi 0}$, we can estimate $\zeta_x$ through
\begin{equation}
\zeta_x^2 = \frac{1}{3} \frac{{g_x}_1^2}{\Omega_{\phi 0} (1+w_0)} \,,\label{zeta1}
\end{equation}
which is nothing more than Eq.~(\ref{eqw}) at present (noting that $1+1/\sigma a^3 = 1/\Omega_\phi$). Once this coupling has been estimated we can proceed by solving Eq.~(\ref{eqsigma}) therefore obtaining the evolution of the equation of state at large redshifts only using information based on the variation of constants. In practice current observations only constrain effective values of $\Omega_{\phi}$ and $w$ (a weighted average over redshift) rather that their present day values. However, our simpler assumption will not greatly affect our results. We also stress that for any given $\zeta_x$ it is possible to obtain any evolution for $\Delta x / x$ by choosing appropriate scalar field potential and initial conditions. This means that $\zeta_x$ can only be effectively constrained using complementary datasets.

If the same dynamical scalar field providing the dark energy is also responsible for the varying couplings there will be an important consistency test provided by violations of the Equivalence principle. Indeed, the smaller $\zeta_x$ is, the larger $\phi$ needs to be in order to be responsible for a given evolution of $\Delta x/x$. Hence, $\zeta_x$ cannot be made arbitrarily small or otherwise the kinetic energy of the field would be too large for it to be the dark energy. In this context, it has been shown \cite{Damour1,Damour2,Oliveira} that, if at least some of the claimed detections of varying couplings are correct then there must be violations of the Einstein Equivalence Principle at a level within reach of the forthcoming generation of space-based experiments (ACES, $\mu$SCOPE, GG and STEP), which is expected to improve on the present sensitivity by as much as five orders of magnitude. Hence, one can envisage that these experiments might themselves soon provide a measurement of $\zeta_x$ rather than just bounds.

There is also the possibility of combining our reconstruction with a standard method such as using type Ia Supernovae. Despite its weaknesses, which we discussed above, reconstruction using type Ia Supernovae is a more direct probe of the evolution of $\phi$ and will provide an important consistency check 
\cite{Nakamura:1998mt,Saini:1999ba,Chiba:2000im,Alam:2003sc,Alam:2003fg,Zhang}. In particular, we should be able to test, though with limited accuracy, the validity of our main assumption (the linear relation between $\Delta x / x$ and $\phi$ given by Eq.~(\ref{lin})) using future space-based experiments such as DUNE or JDEM.

\section{Simulated data}

We perform an analysis having in mind the quality of the datasets expected
to be available both in the near future (say in 2 or 3 years time) and what
is expected for the next generation of ground-based astronomical
instrumentation. Specifically, for the latter case we focus on the expected
sensitivity of the CODEX instrument for the European ELT (formerly known as
OWL).

Note that we will restrict our attention to direct measurements of the various couplings. The analysis of the Oklo natural nuclear reactor (see \cite{Petrov} for a critical review) offers very accurate measurements of $\alpha$ at $z\sim0.14$, $\Delta \alpha / \alpha \sim 10^{-7}$. Furthermore this limit is strengthened by about two orders of magnitude if one assumes that other couplings are varying \cite{Olive}. Nevertheless, these methods rely on data (and theory) beyond our direct control and as such will not be applied here. Another example of an indirect measurement is one based on
OH lines in gravitational lens systems at a redshift of order $z \sim 0.7$
 \cite{Kanekar}. On the other hand, measurements using the CMB \cite{cmb1,Martins,Rocha,Ichikawa} are direct but share with Oklo and also with local laboratory experiments \cite{Marion,Peik} the disadvantage of measuring $\alpha$ mainly at a single redshift, which makes the inter-comparisons subtle (due to the differing systematics), as well as model-dependent. Methods that can probe a wide range of redshifts are therefore best suited for our purposes.

We produced simulated data sets for both $\alpha$ and $\mu$ using a Monte Carlo generator based on the redshift dependence of the scalar field for three  particular scalar potentials under a corresponding choice of the parameter $\zeta_x$.
The evolution of the field with redshift was determined by numerically integrating the field's equation of motion for the chosen potentials. 
These are models previously investigated in the literature and we fix their corresponding parameters so that the various examples can be representative of the possible evolutions of the equation of state. Two of the models are given by a potential with two exponential terms \cite{Barreiro}. Model A, is given by 
$V(\phi) = V_0(\exp(10\kappa \phi) + \exp(0.1 \kappa \phi))$ and model B by
$V(\phi) = V_0(\exp(50\kappa \phi) + \exp(0.8 \kappa \phi))$.
The first term allows the energy density of the field to approach an attractor solution for a wide range of initial conditions and the second term drives the Universe into a period of accelerated expansion in accordance with current observational constraints. In the first model the equation of state parameter $w$, increases with redshift and in the second it decreases.
The third model, model C, is given by, 
$V(\phi) = V_0 \exp(2 (\kappa \phi)^2)$ \cite{Lee2}. For large values of the scalar field, this potential also has an attractor solution with $w \approx$ constant, however,  the equation of state parameter quickly decreases to $-1$ as the field approaches the minimum.

Based on the current data and the theoretical estimates for the parameter $R$, we consider a fiducial model in which the parameter $\zeta_x$ is fixed is such a manner that we obtain $(\Delta \alpha/\alpha)_{z=3} = -0.06 \times 10^{-5}$ and $(\Delta \mu/\mu)_{z=3} =  3 \times 10^{-5}$, thus $R = -50$.
For models A, B and C, it can be found that 
the appropriate values are $\zeta_\alpha = 1.6 \times 10^{-6}$, $\zeta_\alpha = 2.07 \times 10^{-6}$ and $\zeta_\alpha = -0.74 \times 10^{-6}$, respectively.
 
The distributions of the data sets are equivalent to the current Murphy {\it et al.} data \cite{Murphy}. This data spans the interval $z = [0.2,3.7]$, however, we interpolate the distributions up to redshift $z=4.2$ as a damped Lyman-$\alpha$ (DLA) absorption system has been recently identified at this redshift. 
 
\begin{figure}[!ht]
\includegraphics[width = 8.5cm]{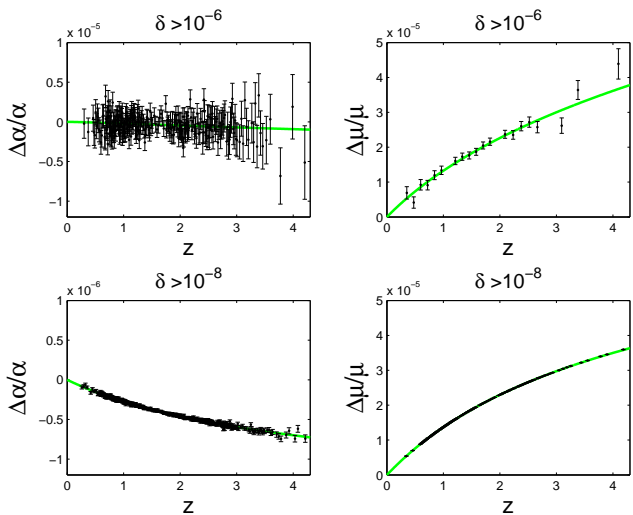}
\caption{\label{data} Potential data sets, generated from model A, expected in the relatively near future (top two panels labeled with $\delta > 10^{-6}$) and with CODEX (lower two panels labeled with $\delta > 10^{-8}$) as described in the text are compared. $\delta$ represents the minimum value of the error bars considered in the corresponding simulated data.}
\end{figure}

We assume that in the near future we will be able to determine the variation in $\alpha$ from a sample of 250 objects and the variation in $\mu$ from 20 objects. We take the minimum value of the error bars to be $\approx 10^{-6}$ occurring at a redshift $z \approx 1.5$ and increasing quadratically from this redshift. The coefficient of this quadratic function and the redshift of minimum error bar were also determined from the present data set. An example of the typical simulated data sets is shown in the
upper two panels of Fig.~\ref{data}. For the CODEX spectrograph the precision in $\alpha$ is estimated to be $\approx 10^{-8}$. We use this value for the minimum error bar in the simulated data. As before, the error bars of the remaining data follow a quadratic function with minimum near redshift $z = 1.5$. We conservatively use a sample of 500 data points for $\alpha$ and of 100 data points for $\mu$. The high
resolution simulated data is shown in the lower two panels of Fig.~\ref{data}. 

\section{Results}
We have performed a likelihood analysis by fitting the various parameters ${g_x}_i$ to the simulated data samples. When using both samples together, we fit ${g_\mu}_i$ and in addition $R$ such that the full vector sample is $[R (\Delta \alpha/\alpha)_{j_\alpha}, (\Delta \mu /\mu)_{j_\mu}]$, where $j_\alpha$ and $j_\mu$ are indices running over the size of the $\alpha$ and $\mu$ samples, respectively.

\subsection{Near-future prospects}
\begin{figure}[!ht]
\includegraphics[width = 8.5cm]{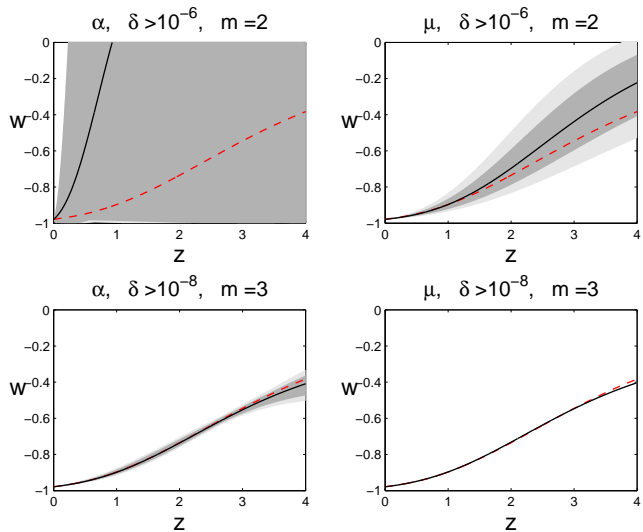}
\caption{\label{recw1} The reconstruction to the $m$th order of the equation of state and its error band is shown for the various surveys assuming the potential $V(\phi) = V_0(\exp(10 \kappa \phi) + \exp(0.1 \kappa \phi))$ (model A). The dashed line represents the dark energy equation of state corresponding to the potential used to generate the simulated data and the solid line corresponds to the reconstruction's best fit. The dark region is the $1\sigma$ confidence level on the parameters and the light region the $2\sigma$ confidence level.}
\end{figure}

\begin{figure}[!ht]
\includegraphics[width = 8.5cm]{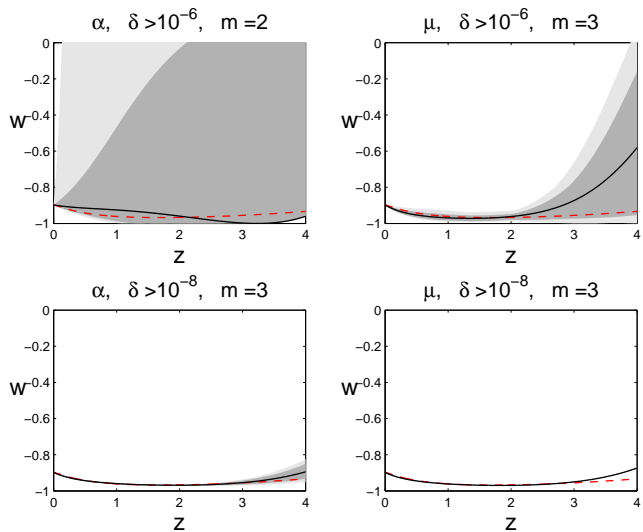}
\caption{\label{recw2} Same as in Fig.~\ref{recw1} with the potential $V(\phi) = V_0(\exp(50 \kappa \phi) + \exp(0.8 \kappa \phi))$ (model B).}
\end{figure}

\begin{figure}[!ht]
\includegraphics[width = 8.5cm]{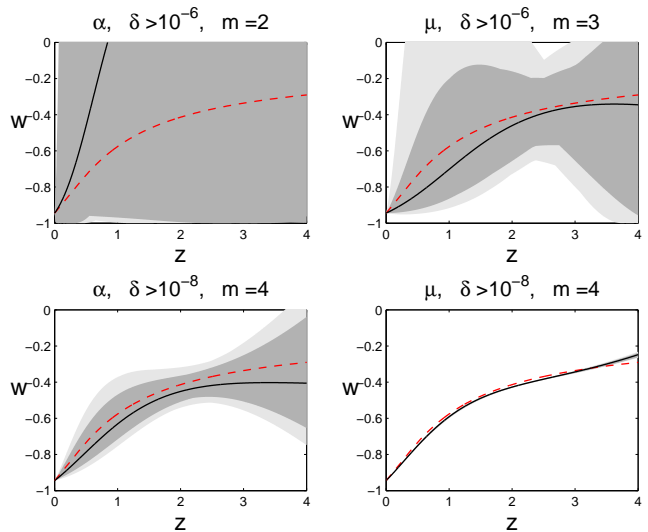}
\caption{\label{recw3} Same as in Figs.~\ref{recw1} and \ref{recw2} with the potential 
$V(\phi) = V_0 \exp\left(2 (\kappa \phi\right)^2)$ (model C).}
\end{figure}

In Figs.~\ref{recw1}, \ref{recw2} and \ref{recw3} 
we show the potential for reconstructing the equation of state parameter, $w$ as a function of redshift.
The dark (light) shaded regions correspond to the 1 (2) $\sigma$ confidence levels of reconstruction.
We clearly observe from the upper left panels of Figs.~\ref{recw1}, 
\ref{recw2} and \ref{recw3} that the reconstruction based entirely on $\Delta \alpha/\alpha$ will still be unsatisfactory in the near future. The reconstruction shown in the upper right panel of Figs.~\ref{recw1}, \ref{recw2} and \ref{recw3} of the equation of state based on the $\mu$ sample, however, will be very illuminating. For the fiducial model under study, 20 points are quite sufficient to make it clear that $w$ increases with redshift. Given that the signal in $\mu$ is fifty times larger than the one on alpha, it is no surprise that the reconstruction is in this case more successful.  We can read from Tables~\ref{tab1}, \ref{tab2} and \ref{tab3} that we can obtain an improvement on the reduced $\chi^2$ going to second order in the degree of the polynomial $g_\mu$, though a third parameter is only helpful when reconstructing model B. Because the data on $\alpha$ is comparatively so poor, when combining both samples we obviously find that the parameter $R$ is also poorly constrained (see Fig.~\ref{liker}).


\begin{table}
\begin{tabular}{cc|cc|cc}
\hline
~ & ~ & Near future & ~ & CODEX & ~  \\
~ & $m$ & $\chi^2/{\rm dof}$ & $\Delta \chi^2$ & $\chi^2/{\rm dof}$ & $\Delta \chi^2$ \\
\hline
~ & 1 & 0.863 & ~ &  1.670 & ~ \\
$\alpha$  & 2 & 0.865 & $0.4$ &  0.936  & $367$ \\
~ & 3 & ~ & ~ & 0.923 & $7.4$ \\
\hline
~ & 1 & 1.275 & ~ & $2\times 10^3$ & ~ \\
$\mu$  & 2 & 0.958 & $7.0$ & 39 & $2 \times 10^5$  \\
~ & 3 & 1.007 & $0.1$ & 1.2 & $3.7 \times 10^3$  \\
\hline
~ & 1+1 & 0.892 & ~ & 327 & ~ \\
~ all ~ & 2+1 & 0.869 & $7.2$ &  7.1 & $1.9 \times 10^5$ \\
~ & 3+1 & 0.872 & $0.01$ &  0.96 & $3.7 \times 10^3$ \\
\hline
\end{tabular}
\caption{\label{tab1} Best fit results when fitting the $\alpha$ samples and $\mu$ samples with $m$ ${g_x}_i$ parameters, and both samples with the additional parameter $R$. The left columns shows the reduced $\chi^2$ and $\Delta \chi^2 = \chi^2(m-1)-\chi^2(m)$, for the type of data expected in the near future and the right columns the same quantities for data expected with CODEX. The data was generated from the evolution using the potential 
$V(\phi) = V_0(\exp(10 \kappa \phi) + \exp(0.1 \kappa \phi))$ (model A).}
\end{table}
\begin{table}
\begin{tabular}{cc|cc|cc}
\hline
~ & ~ & Near future & ~ &  CODEX & ~  \\
~ & $m$ & $\chi^2/{\rm dof}$ & $\Delta \chi^2$ & $\chi^2/{\rm dof}$  
& $\Delta \chi^2$ \\
\hline
~ & 1 & 0.969 & ~  &  29 & ~ \\
$\alpha$  & 2 & 0.970 & $0.9$ &  1.359 &  $1.4 \times 10^4$ \\
~ & 3 & ~ & ~ &  0.948 & 206  \\
\hline
~ & 1 & 7.6 & ~ & $7.2 \times 10^4$ & ~ \\
$\mu$  & 2 & 1.5 & $117$ & $1.3 \times 10^3$ & $7 \times 10^6$ \\
~ & 3 & 1.12 & $7.9$ & 7.1  & $1.3 \times 10^5$  \\
\hline
~ & 1+1 & 1.439 & ~ & $1.2 \times 10^4$ \\
~ all ~ & 2+1 & 1.001 & $118$ &  217 & $7 \times 10^6$  \\
~ & 3+1 & 0.977 & $7.4$ & 1.95  & $1.3 \times 10^5$  \\
\hline
\end{tabular}
\caption{\label{tab2} Same as in Table~\ref{tab1} for the potential 
$V(\phi) = V_0(\exp(50 \kappa \phi) + \exp(0.8 \kappa \phi))$ (model B).}
\end{table}
\begin{table}
\begin{tabular}{cc|cc|cc}
\hline
~ & ~ & Near future &~ &  CODEX & ~  \\
~ & $m$ & $\chi^2/{\rm dof}$ & $\Delta \chi^2$ & $\chi^2/{\rm dof}$ & $\Delta \chi^2$ \\
\hline
~ & 1 & 1.034 & ~  &  1.736 & ~ \\
$\alpha$  & 2 & 1.036 & $0.5$ &  1.233 & $252$  \\
~ & 3 & ~ & ~ &  1.108 & $77$ \\
~ & 4 & ~ & ~ & 1.108 & $1.2$  \\
\hline
~ & 1 & 1.483 & ~ & $1.5 \times 10^3$ & ~ \\
$\mu$  & 2 & 1.349 & $3.9$ & 399  & $1 \times 10^5$ \\
~ & 3 & 1.382 & $0.79$ & 21 & $3.7 \times 10^4$ \\
~ & 4 & ~ & ~ & 1.6 & $1.8 \times 10^3$   \\
\hline
~ & 1+1 & 1.066 & ~ & 245 & ~ \\
~ all ~ & 2+1 & 1.055 & $4.0$ &  67 & $1 \times 10^5$  \\
~ & 3+1 & 1.056 & $0.7$ & 4.2 & $3.7 \times 10^4$ \\
\hline
\end{tabular}
\caption{\label{tab3} Same as in Tables~\ref{tab1} and \ref{tab2} for the potential $V(\phi) = V_0 \exp(2(\kappa \phi)^2)$ (model C).}
\end{table}

Alternatively, one could consider generating a simulation based on the result from the analysis in Ref.~\cite{Murphy} implying that $(\Delta \alpha/\alpha)_{z = 3} \approx -0.5\times 10^{-5}$ (in which case we would have $R\sim -5$).  In this case, we would conclude that the $\alpha$ sample, though not as good as a $\mu$ sample, provides nonetheless a comparable reconstruction with a similar error band and the accuracy of the determination of $R$ can be significantly better than the one presented in Figs.~\ref{recw1}, \ref{recw2} and \ref{recw3}.

\subsection{The promise of CODEX}

\begin{figure}[!ht]
\includegraphics[width = 8.5cm]{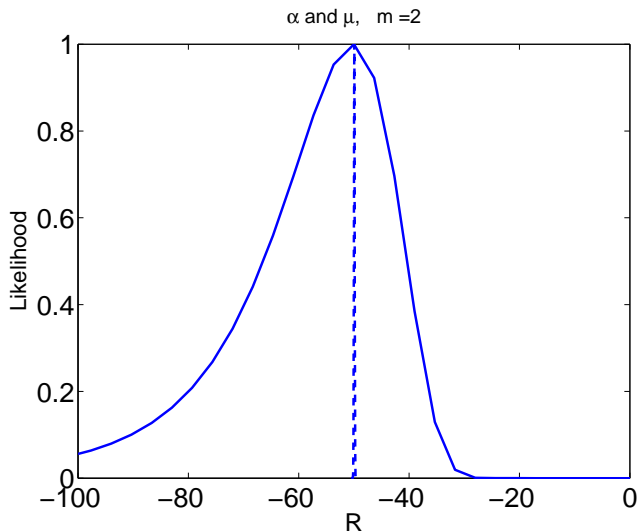}
\caption{\label{liker} The precision achieved on the parameter $R$ with the various surveys by fitting $\alpha$ and $\mu$ samples together generated from model A. The solid line illustrates the possible likelihood to be obtained with near future data and the dashed line, the likelihood using the CODEX data.}
\end{figure}

We now turn our attention to the reconstruction of the equation of state derived from measurements with the CODEX instrument \cite{Codex,Molaro}. In this case, the samples allow for a precise reconstruction and provide the possibility to go beyond the first coefficients of the polynomial.  From Tables~\ref{tab1}, \ref{tab2} and \ref{tab3} we see that a fit for $\alpha$ involving at least 
three parameters is now preferable. In principle the error bars are small enough to admit a polynomial of even higher degree (see lower panels of Figs.~\ref{recw1}, \ref{recw2} and \ref{recw3}. 
This will be in fact necessary if the quintessence field oscillates at the minimum of its potential. There are, however, difficulties associated with models with oscillatory behaviour as explained in Ref.~\cite{Nunes}. These models have typically an equation of state near $-1$ which introduces a large uncertainty in the value of $\zeta_x$ and consequently on the overall reconstruction of the equation of state. However, as discussed above this difficulty might be avoided with future datasets. Combining both data sets, the uncertainty on $R$ will be limited to only a fraction of a percent as can be shown in Fig.~\ref{liker}.

Consider a sample with $N_\alpha$ and $N_\mu$ independent data points for 
$\Delta \alpha/\alpha$ and $\Delta \mu/\mu$ with a 
given redshift distribution. Let us also assume a given redshift 
distribution for the individual $\Delta \alpha/\alpha$ and $\Delta \mu/\mu$ 
error bars with a normalization given respectively by 
$\epsilon_\alpha$ and $\epsilon_\mu$.
The error bands in the reconstruction 
of the equation of state scale approximately as
\begin{equation}
\label{scaling}
\frac{\epsilon_\alpha \epsilon_\mu}{\sqrt{\epsilon_\mu^2 N_\alpha + R^2 \epsilon_\alpha^2 N_\mu}} \,,
\end{equation}
%
where $N_\alpha$ and $N_\mu$ correspond to the size of the $\alpha$ and $\mu$ samples respectively. In fact Eq.~(\ref{scaling}) allows for a 
simple re-scaling of the results if different samples with similar 
redshift distributions for the $\Delta \alpha/\alpha$ and $\Delta \mu/\mu$ 
data points and respective error bars are used. For the simulated data used to 
perform the reconstruction of the equation of state our simple re-scaling 
formula gives a ratio of 14:1:0.1:0.004 for the relative sizes of the 
error bands of the various models considered in Figs.~\ref{recw1}, \ref{recw2} and \ref{recw3} which is roughly in agreement with the above results.

\section{Discussion}

We have discussed the possible use of varying couplings to reconstruct the equation of state of dark energy, and emphasized the high benefit of obtaining further measurements of $\mu$. In combination with measurements of $\alpha$ (which are easier to obtain, given the abundance of sources) they can provide a key test of the relationship between varying couplings in a grand unification scenario. This method not only complements results anticipated by hypothetical future experiments (such as JDEM and DUNE), but given reasonable expectations for forthcoming improvements in spectroscopic measurements, is expected to be competitive with the standard methods for dark energy equation of state reconstruction (both those using supernovae and those based on weak lensing).

Having two different observables (the fine-structure
constant and the proton-to-electron mass ratio), one can in fact check the
self-consistency of the reconstruction. In fact, given that standard
procedures directly probe the evolution of $\phi$ at low redshift, another
consistency check will be provided by combining our equation of state
reconstruction with the standard one. A failure of the consistency check can be interpreted in one of two ways: that the coupling (\ref{lin}) is inadequate and possibly an expansion to higher order in $\phi$ is in order; or that the coupling to matter vanishes altogether and it is impossible to correlate the variations in couplings with dark energy.

For the class of models studied, however, the reconstruction of the dark energy equation of state using varying
couplings has several fundamental advantages over the standard methods.
Firstly, one has the advantage of a much larger lever arm in terms of
redshift, since such measurements can easily be made up to redshifts of
$z\sim 4$. This may not seem like a substantial advantage, since dark energy is
only dynamically relevant at relatively low redshift, but in fact it is a
key one, since one can probe the redshift range where the field evolution
is expected to be fastest (if it is a tracking field)---that is, deep in
the matter era. Secondly, the varying couplings method is also much cheaper and
can be done without any problems from the ground, even with existing
facilities -- all it takes is a few hundred good nights of telescope time,
which is quite a modest investment given the potential gains. Note that the
optimal observing strategy (e.g., the choice of the number of sources to
observe and their redshift distribution), for a given total available
observation time, will depend on a range of factors, both theoretical and
observational. The design of such a strategy is an important issue that
needs to be studied in more detail in the future.

To conclude, let us again stress the power and simplicity of this method.
It requires only ground-based observations, and a relatively small dataset
(possibly requiring not more than a hundred hours of telescope time) can
have a huge impact and conceivably provide unambiguous evidence for
dynamical dark energy. The main point to keep in mind is that a good and
uniform redshift coverage is important, since one will have to calculate
derivatives of the processed data. The small but representative range of
models we have considered shows that this is within reach in the next few
years. Last but not least, this is also an example of how astrophysical
observations can be optimal probes of fundamental physics. We believe that
such astrophysical probes will become increasingly common in years to come,
and hope this provides early encouragement for the observational
astrophysics community.

\section*{Acknowledgments}

This work was funded in part by FCT (Portugal), in the framework of the POCI2010 program, supported by FEDER. Specific funding came from grant POCI/CTE-AST/60808/2004. The work of NJN and KAO was supported in part by DOE grant DE--FG02--94ER--40823.

\bibliography{alphamu}

\end{document}